\def\breakon{\end{multicols}\widetext\vspace{.5cm}
\noindent\rule{.48\linewidth}{.3mm}\rule{.3mm}{.5cm}\vspace{.5cm}}
\def\breakoff{\vspace{.5cm}
\noindent
\rule{.52\linewidth}{.0mm}\rule[-.47cm]{.3mm}{.5cm}\rule{.48\linewidth}{.3mm}
\vspace{.5cm}
\begin{multicols}{2}
\narrowtext}
\documentstyle[aps,epsf,multicol,eqsecnum]{revtex}
\begin{document}
\draft
\widetext
\title{Universal structure of the edge states of the fractional
quantum Hall states}
\author{Ana Lopez$^1$ and Eduardo Fradkin$^2$}
\address{$^1$Centro At{\'o}mico Bariloche,  (8400) S.\ C.\ de Bariloche,
R{\'\i}o Negro, Argentina\\
$^2$Department of Physics, University of Illinois at
Urbana-\-Champaign, 1110 West Green Street,Urbana, IL 61801-3080}
\bigskip
\maketitle

\begin{abstract}
We present an effective theory for the bulk fractional quantum
Hall states on the Jain sequences on closed surfaces and show that
it has a universal form whose structure does not change from
fraction to fraction. The structure of this effective theory
follows from the condition of global consistency of the flux
attachment transformation on closed surfaces. We derive the theory
of the edge states on a disk that follows naturally from this
globally consistent theory on a torus. We find that, for a fully
polarized two-dimensional electron gas, the edge states for all
the Jain filling fractions $\nu=p/(2np+1)$ have only one
propagating edge field that carries both energy and charge, and
two non-propagating edge fields of topological origin that are
responsible for the statistics of the excitations. Explicit
results are derived for the electron and quasiparticle operators
and for their propagators at the edge. We show that these
operators create states with the correct charge and statistics. It
is found that the tunneling density of states for all the Jain
states scales with frequency as $|\omega|^{(1-\nu)/\nu}$.
\end{abstract}
\bigskip

\begin{multicols}{2}
\narrowtext

\section{Introduction}
\label{sec:intro}

The properties of the fractional quantum Hall states have been
studied extensively, beginning with Laughlin's microscopic
theory\cite{laughlin} and its
generalizations\cite{haldane-h,halperin-h,jain-h}, followed by
Chern-Simons field theory approaches\cite{zhk,paper1}, and ending
with a classification of all possible abelian FQH states in the
form of effective low energy theories that capture their universal
features\cite{blok-wen,wenzee-matrix,wen,ga}. Much of the present
understanding of the universal properties of the abelian FQH
states is encoded in the effective lagrangian of Wen and
Zee\cite{wenzee-matrix}which is given by
\begin{eqnarray}
{\cal L}_{\rm eff} =&& \sum_{I,I'} {1\over 4\pi} \; K_{II'} \;
\epsilon_{\mu\nu\lambda}\;  a_{I\mu} \;  \partial_{\nu} \;
a_{{I'}\lambda} -\sum_I\ell_I j_\mu^v a_I^\mu
\nonumber \\
&& -\sum_I{1\over 2\pi} \;
\epsilon_{\mu\nu\lambda}\; A_{\mu}\; \partial_{\nu} \; t_{I} \;
a_{I\lambda}
 \label{eq:a5}
\end{eqnarray}
where, for the Jain  states, {\it i.\ e.\/}, for states with $\nu=
{p \over {2np+1}}$, the matrix $K$ and the charge vector $t$ take
the values
\begin{equation}
K= 1_{p \times p} + 2n C , \; t^{T}=(1,....,1)
\label{eq:a9}
\end{equation}
where $C$ is a $p \times p$ matrix with all its elements equal to
one. In Eq.~( \ref{eq:a5}) $j_\mu^v$ is the vortex current,
$\ell_I$ is the vector of vortex (quasiparticle) charges and $t_I$
is the electromagnetic charge vector. A key topological property
of the abelian QH fluids is the topological degeneracy of the
Hilbert space (not just the ground state) which is equal to $|\det
K|^g$, where $g$ is the genus ({\it i.\ e.\/}, the number of
handles) of the surface (manifold) on which the fluid
moves\cite{wen-niu}. Wen\cite{wen} has emphasized the concept of
topological order of the fluid and characterized it by the
degeneracy of the Hilbert space.

This general classification of the abelian FQH states is very
powerful. In addition of giving a compact and universal
description of all abelian FQH states, it also classifies the
possible behaviors of their associated edge
states\cite{wen,wen-edge,stone-edge}. The central point of this
classification is that for a {\it given} effective theory of the
bulk incompressible state there is a {\it natural} construction of
the edge in terms of the conformal field theory of chiral
bosons\cite{wen-edge}. A fundamental feature of this construction
is the existence of a one-to-one correspondence between the
quasiparticle states of the bulk and the primary fields that build
the spectrum of the edge states. This construction assumes that
the behavior of the electron gas near the edge is essentially
simple, which may or may not be realized if edge reconstruction
actually takes place\cite{brey}. Whether or not this happens
depends on many non-universal microscopic issues, such as edge
potentials, interactions and impurities, which complicate matters
and which may be quite relevant for a precise understanding of the
edge tunneling experiments away from the middle of the bulk Hall
plateaus\cite{chang}. Thus, in realistic situations, the detailed
microscopic physics near the edge may actually mask the robust
physics of the bulk. However, once the details of edge
reconstruction are sorted out, it is expected that the physics of
the edge should be universal even though its connection with the
physics of the bulk is no longer so simple.

Microscopically, the bulk states can be constructed by
implementing the idea of flux-attachment, by coupling particle
currents to  a suitable set of Chern-Simons gauge
fields\cite{zhk,paper1}. These ideas, which are at the root of the
construction of fractional (or braid) statistics\cite{wz,wuzee},
play a central role in the field-theoretic descriptions of the
fractional quantum Hall effect for both single layer
systems\cite{zhk,paper1,book,capitulo,zhang,hlr},
bilayers\cite{wzphonon,ezawa,paper2} and for spin-singlet or
partially polarized FQH states \cite{paper2,FB}. Since the
effective action of any incompressible state of a system of
charged particles in two dimensions in the presence of a strong
magnetic field, by general hydrodynamic
arguments\cite{frohlich-zee,frohlich-kerler}, {\it must} be of the
Chern-Simons form, the effective action that is actually {\it
derived} from the field theoretic descriptions is a Chern-Simons
gauge theory and it fits the $K$-matrix classification.

However, a number of puzzles implicit in this picture have not
been fully resolved. In particular, the $K$-matrix classification
assumes that the number of physically distinct stable
quasiparticles is equal to the rank of the matrix. This feature is
quite puzzling since the only known conservation law in this
system is just charge conservation and it is unclear from where
the necessary additional conservation laws may come from. Also,
this same problem is connected with two important features of the
$K$-matrix classification. One is the fact that for a given
filling fraction there are many possible physically distinct
theories, characterized by different charges and statistics of the
quasiparticles. It has been argued by Wen that these states have
topological orders that tell these different states apart. For
example, the $\nu=2/5$ single layer, fully polarized state is
given by a $2 \times 2$ $K$ matrix  and it seems to be closely
related to the $\nu=2/5$ fully polarized QH state in bilayers.
~From a hydrodynamic point of view, in the absence of interlayer
tunneling, it is obvious that the bilayer system should have two
conserved currents and hence a $2 \times 2$ $K$ matrix. However,
the physical meaning of these two separately conserved
hydrodynamic currents {\sl in single layer systems} is unclear
since there is no known conservation law to insure their
stability. A closely related problem is the associated composite
picture of the edge states with a number of branches equal to the
rank of the $K$-matrix. In practice, these structures have their
origin in the multiple condensates of the hierarchical
construction of the FQH states. In contrast, in the fermion
Chern-Simons theory description of the 2DEG on a disk, which
should be equivalent to the $K$-matrix picture, there is only one
gauge field and only one quasiparticle whereas the $K$ matrix
classification typically will have several gauge fields and the
associated quasiparticles. It is natural to ask what is the
minimal universal structure  required to describe the abelian FQH
states in the bulk. Similar puzzles arise in the description of
the edge states, which require a composite structure (even in the
absence of edge reconstruction) with a quite complex behavior as a
function of the filling factor. In particular, the contribution of
the edge states to the specific heat of the 2DEG (in fact, the
total contribution!) looks like a number-theoretic function which
diverges at the compressible fractions. However, this picture
holds only if the currents that define the structure of the $K$
matrix for the composite edges are actually conserved.

In addition, the standard mean field theory constructions of the
bulk states are actually inconsistent for a system on a closed
surface such as a torus. This is an important problem since the
degeneracy of the Hilbert space on a closed surface is a universal
feature of these topological fluids and it is closely related to
the puzzles mentioned above. The usual flux-attachment
transformation \cite{zhk,paper1,hlr} is  regarded as a process in
which a particle is physically and {\it locally} glued to a
certain number of (statistical) flux quanta. For an electron gas
on an {\it open} simply connected disk with fixed boundary
conditions, there is no problem with this procedure. However, if
we were to carry out this procedure on a closed surface, such as a
torus or a sphere, this approach  is not consistent as it is
usually described. Similar difficulties arise if the disk is
replaced by an annulus. The problem is the role of gauge
transformations that wind around non-contractible loops on the
surface. This issue is important since it is precisely these gauge
transformations that carry the information on the topological
order ( {\it i.\ e.\/} degeneracy), and hence they provide an
essential consistency condition.

In this paper we construct a theory of the FQH fluids for a single
layer, fully polarized 2DEG, with filling fraction on a Jain
state, on a closed surface. We derive the effective
Chern-Simons action of these incompressible  states. To do so it
is necessary to modify the flux-attachement procedure so as to
make it globally consistent.  The key issue is the invariance (or
lack of) of the Chern-Simons action under large gauge
transformations for a theory defined on a closed manifold with a
number of handles $g$. (In practice we will only be interested on
describing the behavior of the system on a torus, and so we will
use $g=1$.) It has been shown by a number of
authors\cite{witten,hosotani1,hosotani2} that invariance under
large gauge transformations of the path-integral (not the action!)
requires that the coupling constant of any Chern-Simons theory
(both abelian and non-abelian) {\it must} be quantized. However,
the standard Chern-Simons constructions of the flux-attachment
transformation violate this principle. Here we make use of the
results of reference\cite{chetan} to derive an effective theory
that is globally consistent. We will show that the resulting
effective theory can indeed be cast into a $K$-matrix form but
with a different and much simpler structure than the usual one. We
find that this structure requires only a small number of gauge
fields and their number, {\it i.\ e.\/} the rank of the $K$ matrix, is
fixed and independent of the filling fraction. We find that for each
Jain state there is only one quasiparticle. Furthermore, we also show
that the consistent theory on a closed surface has a related unique
universal edge structure on an open surface. This edge structure is
particularly simple.

As this manuscript was being prepared we became aware of a very
recent unpublished work by D.\ H.\ Lee and X.\ G.\ Wen
\cite{wenlee} where they argued that the effective theory of the
edge states for the Jain fractions contains just two independent
chiral bosons, one of which does not propagate. Our final result
in essence agrees with the picture advocated by Lee and Wen.
However, the underlying philosophy is quite different since in our
work the non-propagating fields have a purely topological origin.

This paper is organized as follows. In section \ref{sec:topology}
we introduce a general framework for flux attachment that is
consistent on closed manifolds such as a torus. Here we show that
the effective theory of the FQH states for $\nu={\frac{p}{2np \pm
1}}$, has a simple and universal structure. In section
\ref{sec:edge} we derive the theory of the edge states that
follows from this universal structure. In section \ref{sec:eops}
we derive the form of the electron and quasiparticle operators at
the edge and compute their propagators. Section \ref{sec:concl} is
devoted to the conclusions.

\section{Flux attachment on Closed Surfaces}
\label{sec:topology}

In ref.\cite{chetan} it was
shown that there is a simple and direct way to reformulate the Chern-Simons
theory of the (single layer, fully polarized) FQH state in order
to satisfy the requirement of global consistency on a closed surface. For a
fermion representation of
the FQH system ({\it i.\ e.\/} composite fermions) it was shown
that the exact partition function can be written as a path integral of a
theory in which the particles whose worldlines are represented by the
currents
$j_\mu$ , interact with two gauge fields $a_\mu$ and $b_\mu$. These
interactions are encoded in the following effective action
\begin{equation}
S_{\rm eff}[a,b,j]= {\frac{1}{2\pi}} a^\mu
\epsilon_{\mu \nu \lambda} \partial^\nu b^\lambda- a^\mu
j_\mu - {\frac{2n}{4\pi}} \epsilon_{\mu \nu \lambda}
b^\mu \partial^\nu b^\lambda
\label{eq:effective}
\end{equation}
Therefore, the amplitudes can be written in terms of a path
integral over an abelian Chern-Simons gauge field with a correctly
quantized coupling constant equal to ${\frac{2n}{4\pi}}$. Hence
there exists an exact rewriting of the theory involving two gauge
fields, $a_\mu$ and $b_\mu$. These two gauge fields arise quite
naturally: the field $b_\mu$ arises from the fact that the
particle currents (worldlines) are conserved and the field $a_\mu$
is the Lagrange multiplier that imposes the
hydrodynamic constraint between the current and the curl of
$b_\mu$.

The usual form of the flux-attachment transformation is found by integrating
out the gauge field $b_\mu$. For vanishing boundary conditions at infinity,
this leads to an effective action for the field $a_\mu$ of the conventional
form~\cite{paper1}
\begin{equation}
S_{\rm eff}[a]={\frac{1}{4\pi 2n}} \int d^3x
\epsilon_{\mu \nu \lambda} a^\mu \partial^\nu a^\lambda
\label{Sa}
\end{equation}
However, this form of the effective action is not valid for manifolds with
non-trivial topology. Nevertheless, Eq.~(\ref{eq:effective}) is correct
in all cases as it is invariant under both local and large gauge
transformations.

As usual\cite{paper1}, the mean field theory in the composite fermion
language proceeds by first spreading out the field and constructing an
effective integer Hall effect of the partially screened magnetic field. The
result is a description of the states in the generalized Jain hierarchies
$\nu_\pm(n,p)= {\frac{p}{2np\pm1}}$, where $p,n \in {\bf Z}$ and $\pm$ stands
for an electron and hole-like FQH state respectively.

The effective action in the composite fermion picture is found by integrating
out the local particle-hole fluctuations of the fermions about the uniform
mean field state. At long distances and low energies the effective Lagrangian
once again involves a $2 \times 2$ $K$-matrix and it has the usual
Wen-Zee form
\begin{equation}
{\cal L}_{\rm eff}={\frac{1}{4\pi}}K^{IJ} \epsilon^{\mu \nu\lambda}
a_\mu^I \partial_\nu a_\lambda^J
\label{eq:FCS-WZ}
\end{equation}
with
\begin{equation}
K^{IJ}=
\left(
\begin{array}{cc}
\pm p &   1 \\
    1 & -2n
\end{array}
\right)
\label{eq:FCS-Keff}
\end{equation}
We now notice that this effective theory is globally well defined since the
Chern-Simons coupling constants are correctly quantized. Indeed, if we
integrate out the gauge field $b_\mu=a^2_\mu$, we find the
same effective action for $a_\mu$ of our previous work\cite{paper1}. Since
the absolute value of the determinant is $|\det K|=|2np\pm1|$, we find that
the generalized Jain states are $|2np \pm 1|$-fold degenerate on the torus,
which is the correct result.

In what follows we will consider the effective Lagrangian of Eq.~
(\ref{eq:FCS-Keff}) expanded to include the the quantum dynamics
of the quasiparticles. The effective Lagrangian now reads
\begin{eqnarray}
{\cal L}_{\rm eff} = &&{\frac{ p}{4\pi}}\epsilon_{\mu\nu\lambda}
a_{\mu} \partial_{\nu} a_{\lambda}  +{\frac{1 }{2 \pi}}
\epsilon_{\mu\nu\lambda} a_{\mu} \partial_{\nu} b_{\lambda}
\nonumber\\
&&
 - {\frac{2n }{4 \pi}}\epsilon_{\mu\nu\lambda} b_{\mu}
\partial_{\nu} b_{\lambda}   +
{\frac{ 1}{4\pi}}\epsilon_{\mu\nu\lambda} e_{\mu} \partial_{\nu}
e_{\lambda} \nonumber\\ && +{\frac{1}{2 \pi}}
\epsilon_{\mu\nu\lambda} b_{\mu} \partial_{\nu} A_{\lambda} -
(a_\mu+e_\mu) j^\mu_{qp}
 \nonumber\\ && \label{eq:a1'}
\end{eqnarray}
The current $j^\mu_{qp}$ in the last term of Eq.\ (\ref{eq:a1'})
represents the effects of quasiparticles. However, in the fermionic picture
the bare quasiparticles are
composite {\it fermions} whose statistics is modified by the
Chern-Simons gauge fields. Thus, the statistics of all excitations
that we will compute is defined relative to {\it fermions}. A
simple way to keep track of the underlying statistics is to
introduce, as we did in Eq.\ (\ref{eq:a1'}), an additional
Chern-Simons gauge field $e_\mu$ which couples only to the
quasiparticle current $j^\mu_{qp}$.
~From now on, and in order to simplify the notation, we shall call
$p= \pm p$. This
 effective Lagrangian, includes the coupling to a weak external gauge field
$A_\mu$.

We can write the effective lagrangian in a more compact form if we define
$a^1_{\mu}=b_{\mu}$, $a^2_{\mu}=a_{\mu}$ , $a^3_\mu=e_\mu$, the charge vector
$t_I=(1,0,0)$ and the flux vector
$\ell_I=(0,1,-1)$, as
\begin{equation}
{\cal L}= {\frac{1}{4\pi}}K_{IJ}\epsilon_{\mu \nu \lambda}
a_I^\mu\partial^\nu a_J^\lambda +{\frac{1}{2\pi}} t_I\epsilon_{\mu
\nu \lambda} a_I^\mu\partial^\nu A^\lambda + \ell_I a_I^\mu
j^{\mu}_{qp} \label{eq:eff9}
\end{equation}
where the coupling constant matrix is
\begin{equation}
K_{IJ}=\left(
\begin{array}{ccc}
  -2n &    1   &  0  \\
    1 &    p   &  0  \\
    0 &    0   &  1  \\
\end{array}
\right)
\label{eq:Kmat}
\end{equation}
whose determinant is $|\det K|=2np+1$. Hence, we get the correct
degeneracy on closed surfaces.

Following Wen \cite{wen} we can compute  the filling fraction which is given by
\begin{equation}
\nu=| t^{T} K^{-1} t| =  {\frac{p}{2np+1}}
\end{equation}
The quantum numbers of the quasiparticles are
\begin{eqnarray}
     Q_{qp}=&& -e t^{T} K^{-1} l = {\frac{-e}{2np+1}}
\\
{\frac{\theta_{qp}}{\pi}}=&&   l^{T} K^{-1} l =
{\frac{2n}{2np+1}}+1 \nonumber\\ && \label{eq:qn}
\end{eqnarray}
for the charge and the statistics respectively. For the special
case of the Laughlin states, $p=\pm 1$, the gauge field $a^2_\mu$
can be integrated out and the effective action is now identical to
the dual action found by Wen\cite{wen}.

Therefore, the theory defined by Eq. (\ref {eq:eff9}) gives the
correct quantum numbers for the quasiparticle as well as
the correct Hall conductance. This
bosonic representation gives an alternative effective theory of
the Jain states and it does not involve a hierarchy of
condensates, as in Wen's construction.  This picture also suggests
that the effective theory of the edge states for the FQH states in
the Jain sequence does not necessarily require a composite
structure of the edge states.

\section{Edge Theory for the Jain States}
\label{sec:edge}

In this section we use the effective theory for all the states in
the Jain sequence, derived in the previous section, to extract the
effective theory for the edge states. The effective Lagrangian of
Eq.\ (\ref{eq:eff9}) is globally well defined (on closed
surfaces), yields the correct ground state degeneracy on the torus
as well as excitations with the correct fractional charge and
statistics. However, unlike the standard hierarchical construction
of the effective theory of the Jain states\cite{wenzee-matrix},
for a generic state in the Jain sequence, the effective Lagrangian
of Eq.\ (\ref{eq:eff9}) contains the same number of gauge fields,
for all filling fractions on the Jain sequences, and it can be reduced to
just a single gauge field for the special case of the Laughlin
states. In a sense, the Lagrangian of Eq.\ (\ref{eq:eff9}) is the
{\sl minimal} effective theory. This effective Lagrangian has the
standard form introduced by Wen and Zee\cite{wenzee-matrix}  and,
following the general arguments of Wen\cite{wen,wen-edge}, it is
straightforward to extract a theory for the edge states, which we
do in this section. Clearly, since the effective theory of the
bulk in general contains just three gauge fields, the number of
edges does not grow from one state in the hierarchy to the next.
In particular this implies that the specific heat of the system
does not grow without limit as one goes up in the hierarchy.
Consequently, the changes in the thermodynamic properties of the
system that occur as the system becomes compressible is not due to
a proliferation of edges but to a physical collapse of the gap in
the spectrum and the resulting failure to separate the edge from the bulk.

Following Wen's approach\cite{wen}, we derive the theory of the
edge states by noting that the Hilbert space of states of a
Chern-Simons theory on a manifold $\Omega$ (which we take to be a
disk) with a spatial boundary $\partial \Omega \cong S_1$ (where
$S_1$ is a circle) has support at the boundary. This is a special
case of a general result originally derived by
Witten\cite{witten}. In principle this is done as follows. One
imagines that there is a (sharp) potential that confines the
electrons on a simply connected region on
the torus isomorphic to a disk. The gauge fields on the region forbidden to
the electrons can be integrated out since they decouple. The effective
action of Eq.\ (\ref{eq:eff9}) can now be written in the form
\begin{eqnarray}
S=&&{\frac{1}{4\pi}} K_{IJ} \int_\Omega d^3x \;\;
\partial^j \left(a_J^0\epsilon_{ij}a_I^i\right)
\nonumber\\
+&&{\frac{1}{4\pi}} K_{IJ} \int_\Omega d^3x \;\;
a_I ^0 \epsilon_{ij} \left(\partial^i a_J^j-\partial^j a_I^i\right)
\nonumber\\
-&& {\frac{1}{4\pi}} K_{IJ} \int_\Omega d^3x \;\;
\epsilon_{ij} a_I^i \partial^0  a_J^j
\nonumber\\
+&&{\frac{1}{2\pi}}  \int_\Omega d^3x \;\;
\left(a_0^I {\cal J}^0_I+a_i^I {\cal J}^i_I\right)
\nonumber\\
&&
\label{eq:effaction1}
\end{eqnarray}
where we have used the current ${\cal J}^i_I$ defined by
\begin{equation}
{\cal J}^\mu_I \equiv t_I \epsilon^{\mu \nu \lambda} \partial_\nu
A_\lambda+2\pi \ell_I j^\mu_{qp}
\label{eq:current}
\end{equation}
We will impose the gauge condition $a_J^0=0$ at the boundary
$\partial \Omega$. In this gauge the first term of Eq.\
(\ref{eq:effaction1}) vanishes. In this form of the action it is
also apparent that the field $a_J^0$  is a Lagrange multiplier
that enforces the local constraint
\begin{equation}
{\cal J}^0_I=-K_{IJ} \epsilon_{ij} \partial^i a_J^j
\label{eq:gauss}
\end{equation}
which is just Gauss' Law. Similarly, the third term of
Eq.\ (\ref{eq:effaction1}) determines the commutation relations.

The solution of Gauss' Law is
\begin{equation}
a_i^I=\partial_i \phi^I
\label{eq:gauss-solution}
\end{equation}
where $\phi^I$ are two multivalued scalar fields, {\it i.\ e.\/}
singular gauge transformations. If the quasiparticles and the external
fluxes are quasistatic bulk perturbations of the condensate, of
quasiparticle number $N_{qp}$ and flux $\Phi=2\pi N_\phi$,
the scalar fields $\phi^I$
at the boundary $\partial \Omega$ must satisfy the conditions
\begin{eqnarray}
\Delta \phi_I= 2\pi \left(K^{-1}\right)_{IJ}
\left[
\begin{array}{c}
N_\phi \\
N_{qp} \\
-N_{qp}
\end{array}
\right]_J
\label{eq:sources}
\end{eqnarray}
where $\Delta \phi_I\equiv \oint_{\partial \Omega}dx_i
\partial_i\phi_I$ is the change of the field $\phi_I$ once around the
boundary $\partial \Omega$. In components we get
\begin{eqnarray}
\Delta \phi_1=&&{\frac{2\pi}{2np+1}}\left(N_{qp}-pN_\phi\right)
\nonumber\\
\Delta \phi_2=&&{\frac{2\pi}{2np+1}}\left(N_\phi+2n N_{qp}\right)
\nonumber\\
\Delta \phi_3=&& -2\pi N_{qp}
\nonumber\\
&&
\label{eq:bcs}
\end{eqnarray}
In particular, if there is just one quasiparticle in the bulk, $N_{qp}=1$, and
no
extra flux, $N_\phi=0$, we get $\Delta \phi_1={\frac{2\pi}{2np+1}}$,
$\Delta \phi_2=2\pi{\frac{2n}{2np+1}}$ and $\Delta \phi_3=-2\pi$. Conversely,
for $N_\phi=1$ and
$N_{qp}=0$, we get instead $\Delta \phi_1=-2\pi \nu$,
$\Delta \phi_2={\frac{2\pi}{2np+1}}$ and $\Delta \phi_3=0$. Likewise, if we add
an electron to
the bulk, $N_{qp}=2np+1$ but no flux $N_\phi=0$, we get $\Delta \phi_1=2\pi$,
$\Delta \phi_2=2\pi 2n$ and $\Delta \phi_3=-2 \pi (2np+1)$. These conditions
will play an important role below.

Once the constraint Eq.\ (\ref{eq:gauss}) is solved, it is immediate to show
that the content of this theory resides at the boundary $\partial
\Omega$. Indeed, the remaining term in the action Eq.\ (\ref{eq:effaction1})
takes the form
\begin{eqnarray}
S=&&- {\frac{1}{4\pi}} K_{IJ} \int_\Omega d^3x \;\;
\epsilon_{ij} a_I^i \partial^0  a_J^j
\nonumber\\
=&&{\frac{1}{4\pi}} K_{IJ}\int dx_0 \oint_{\partial\Omega}dx_i
\partial^i\phi_I \partial^0\phi_J
\nonumber\\
&&
\label{eq:ccr}
\end{eqnarray}
which is  a theory of chiral bosons. However, as emphasized by
Wen\cite{wen}, as they stand these bosons do not propagate. The
reason is that the Chern-Simons gauge theory is  actually a
topological field theory. Thus, in addition to being gauge
invariant, it is independent of the metric of the manifold where
the electrons reside and hence it is also invariant under
arbitrary local diffeomorphisms. In particular this means that the
Hamiltonian of the Chern-Simons theory is zero. Naturally, this is
just the statement that this is an effective theory for the
degrees of freedom below the gap of the incompressible fluid.
There are no local degrees of freedom left in this regime. The
degrees of freedom only ``materialize" at the boundary which, in
addition to breaking gauge invariance, also breaks the topological
invariance. This is physically obvious since the edge states at
the boundary carry energy and their Hamiltonian does not vanish.

There are several possible ways to represent this physics in the
effective theory. One way is to choose the gauge fixing at the
boundary in a manner that  also breaks the topological invariance.
For instance, Wen\cite{wen} chooses the gauge condition $a_0+v
a_1=0$, where $v$ is chosen to be the velocity of non-interacting
electrons at the edge, {\it i.\ e.\/} $v=cE/B$, with $c$  the
speed of light and $E$  the electric field of the confining
potential at the edge. In the context of the construction that we
are pursuing here, only the gauge field $a_\mu^1$ couples to the
electromagnetic field and thus it is the only one that will
represent propagating degrees of freedom, the charge fluctuations
at the edge.

Another option, which we will make here, is to keep the gauge
condition $a_0=0$, which does not break topological invariance,
but to add boundary terms to the effective action to represent the
effect of the propagating modes at the edge. By power counting,
the boundary term with the smallest scaling dimension one can  add
to the effective action has the form
\begin{eqnarray}
S_{\rm boundary}=&&-\int dx_0 \oint_{\partial \Omega} dx_i
{\frac{{\tilde g}}{2}}
 (a_i^1(x))^2
\nonumber \\ =&&-\int dx_0 \oint_{\partial \Omega} dx_i
{\frac{{\tilde g}}{2}} (\partial_i \phi^1(x))^2 \nonumber \\
\label{eq:bt}
\end{eqnarray}
which is a marginal operator. Here ${\tilde g}$ is a coupling
constant whose physical meaning we discuss below. Notice that this
term only affects the field $a_\mu^1$.

Formally, the boundary term is derived as follows. Within the
framework of the fermionic Chern-Simons theory\cite{paper1}, in
addition to the bulk states there are edge states. A realistic
description of these states requires an understanding of the
problem of edge reconstruction. At the level of a Hartree-Fock
approximation for the fermions in the Chern-Simons picture a
theoretical description was given in
references\cite{chklovskii,vignale}. Although it is not clear if
such descriptions are reliable for systems as quantum mechanical
as the 2DEG in the lowest Landau level, it is clear that an
effective edge must exist even if many of the modes predicted by
 the mean-field theory were to be an artifact of the approximation. In
any event  there should be at least one edge mode that will carry
the correct Hall current at the edge. At the level of the mean
field theory these states are fermionic, as they are in the bulk
before the gaussian fluctuations are integrated out\cite{paper1}.
These edge fermionic states will couple to the {\it boundary}
component of the bulk gauge field $a_\mu^1$. We can now proceed to
integrate out the fermions, as we did before. If the fermions
where non-chiral, the result of integrating out the fermions is
equivalent to conventional bosonization. Their contribution to the
effective action is calculated from the determinant of the Dirac
operator coupled to gauge fields. This is a very standard
result\cite{bosonization} and the effective action is
\begin{equation}
S_{\rm edge}=-{\frac{p}{8\pi}} \int dx_0 \oint_{\partial \Omega} dx_1
(a_\mu^1)^2
\label{eq:edge1}
\end{equation}
which holds in the continuum limit, {\it i.\ e.\/} infinite
bandwidth, and for an infinitesimally narrow edge.

Since the edge theory is actually chiral, we need to keep only the
right moving piece of Eq.~( \ref{eq:edge1}). Hence, $S_{\rm edge}$
becomes
\begin{equation}
S_{\rm edge}=-{\frac{1}{8\pi}} \int dx_0 \oint_{\partial \Omega} dx_1 (a_R^1)^2
\label{eq:edge1'}
\end{equation}
where $a_R^1$ is
\begin{equation}
a_R^1={\frac{1}{\sqrt{v}}} a_0^1+{\sqrt{v}} a_1^1 \equiv {\sqrt{v}} a_1^1
\label{eq:aR}
\end{equation}
with $v=eE/B$  the speed of the edge excitations, and we have used
the gauge condition $a_0^1=0$ at the boundary. Hence, we find
\begin{eqnarray}
S_{\rm edge}=&&- {\frac{pv}{4\pi}}\int dx_0 \oint_{\partial \Omega} dx_1
(a_1^1)^2
\nonumber \\
=&&-{\frac{pv}{4\pi}} \int dx_0 \oint_{\partial \Omega} dx_1
\left(\partial_1 \phi^1\right)^2
\label{eq:bt2}
\end{eqnarray}
which has the form of Eq.\ (\ref{eq:bt}) with ${\tilde
g}={\frac{pv}{2\pi}}$.

The electron-electron interaction term becomes
\begin{eqnarray}
S_{\rm int}=&&\int_{\Omega} d^3x \int_{\Omega} d^3x'
{\frac{1}{2}} (\rho(x)-{\bar \rho}) V(x-x') (\rho(x')-{\bar \rho})
\nonumber\\
=&&\int_{\Omega} d^3x \int_{\Omega} d^3x'  {\frac{1}{8\pi^2}} \epsilon_{ij}
\partial^ia_1^j(x)
V(x-x')
\epsilon_{kl}\partial^ka_1^l(x')
\nonumber\\
\equiv &&
\int dx_0 \oint_{\partial\Omega} dx_i dx'_i
{\frac{t_It_J}{8\pi^2}} \partial_i \phi_I(x) V(x-x') \partial_i \phi_J(x')
\nonumber\\
&&
\label{eq:el-el}
\end{eqnarray}
where we have only retained the boundary contribution since the bulk
excitations have a finite (and for present purposes large) energy gap.
Notice that since $t_I=(1,0,0)$ this term of the action only affects the mode
$\phi_1$. Likewise, the interaction with an external potential with
support at the boundary becomes
\begin{eqnarray}
S_{\rm ext}=&& \int d^3x \left( \rho(x)-{\bar \rho}\right) A_0(x)
\nonumber\\
=&& \int dx_0 \oint_{\partial\Omega} dx_i {\frac{t_I}{2\pi}}
\partial_i\phi_I(x) A_0(x)
\nonumber \\
&&
\label{eq:sext}
\end{eqnarray}
and it involves only $\phi_1$.

Thus, as expected, the effective action is in fact a
theory of edge modes, in agreement with Wen's general arguments. The
effective action involves just three chiral bosons $\phi_I$ (with $I=1,2,3$)
and
takes the form
\begin{equation}
S={\frac{1}{4\pi}}\int_{\partial \Omega \times {\bf R}} dx_0 dx_1
\left( K_{IJ}
\partial_1\phi_I \partial_0\phi_J+ U_{IJ}\partial_1\phi_I \partial_1\phi_J
\right)
\label{eq:edge}
\end{equation}
where $U_{IJ}(x-x')=t_I t_J \left(v+ {\frac{1}{2\pi p}}
V(x-x')\right)$, and its only effect is to determine the velocity
of the edge modes. Here we have used that ${\tilde
g}={\frac{pv}{2\pi}}$. Notice that, as it is well known, the
actual velocity of the edge modes is the sum of two terms, one of
which is determined by the interactions. In what follows we will
work with an effective edge velocity $v$ which includes both the
effects of the edge electric field and of the Coulomb
interactions. Implicitly, and for simplicity, we assume here a
short range interaction. In reality a strict $1/r$ Coulomb
interaction gives a well known logarithmic correction to the
dispersion of the excitations and hence it is not just equivalent
to a redefinition of the velocity. However, this is a well
understood phenomenon which does not affect the main physics of
this system and hence we will work with an effective velocity $v$.
Notice that the only mode with a non vanishing velocity is
$\phi_1$ which is the only mode that couples to perturbations due
to an external electromagnetic field. Thus we identify $\phi_1$ as
the {\it charge mode}. The two remaining modes do not propagate.
Their effect is to fix the statistics of the states.

Finally, we need to relate these fields to edge charge density.
The local charge and current density $J_\mu(x)$ in the plane  is given
by
\begin{equation}
J_\mu(x)={\frac{\delta S}{\delta A_\mu}}={\frac{1}{2\pi}}
t_I \epsilon_{\mu \nu \lambda} \partial^\nu a^\lambda_I
\label{eq:curr}
\end{equation}
The edge currents and densities are integrals of the
bulk currents and densities across the edge. Let $\lambda$ be the
physical width of the edge which we will denote by $\Lambda$ and
it is perpendicular to the edge. Here $\lambda \approx \ell_0$,
the magnetic length. For an edge along the direction $x_1$, the
edge density is given by
\begin{eqnarray}
j_0\equiv&&\int_\Lambda dx_2 J_0(x) \nonumber \\
=&&{\frac{1}{2\pi}}\int_\Lambda dx_2 t_I (\partial^1
a^2_I-\partial^2 a^1_I) \nonumber \\
=&&{\frac{\lambda}{2\pi}} \partial_1{\bar a}^2_1-
{\frac{1}{2\pi}} \Delta a_1^1 \nonumber \\ &&
\label{eq:density1}
\end{eqnarray}
where ${\bar a}^2_1$ is the average of the gauge potential $a^2_1$
across the edge and $\Delta a_1^1$ is the difference of the gauge
potential $a_1^1$ across the edge. We will take the width of the
edge to be infinitesimal $\lambda \to 0$, and since the potential
$a^2_J$ is regular at the edge, the first term in Eq.\
(\ref{eq:density1}) vanishes. With a fixed number of electrons and
at fixed total magnetic field, we can also choose the gauge
potentials to vanish outside the system. Thus,
\begin{equation}
\Delta a^1_1=-a^1_1=-\partial_1\phi_1
\label{eq:bc2}
\end{equation}
where $a^1_j$ is measured {\it inside} the system, at the edge. Hence, the edge
charge density becomes
\begin{equation}
j_0={\frac{1}{2\pi}} \partial_1 \phi_1
\label{eq:density2}
\end{equation}
which is the standard result\cite{wen}. It is straightforward to check
that if $N_{qp}$ quasiparticles are added to the bulk at constant
magnetic field ($N_\phi=0$), the edge acquires a charge
\begin{equation}
Q_{\rm edge}=\int dx_1 j_0(x_1)={\frac{N_{qp}}{2np+1}}
\end{equation}
which, as it should, is equal to the extra charge added to the bulk.

In summary, in this picture there are three edge modes, one
propagating mode associated with charge fluctuations and the other
two non-propagating modes associated with the global topological
consistency of flux-attachment. As we will see in the next section
the only effect of these non-propagating topological modes is to
give the correct statistics to the excitations. Since $|\det
K|=2np+1$ this effective theory reproduces the correct topological
degeneracy of the Hilbert space. Notice that from the point of
view of this effective theory there is no particular difference
between the electron-like FQH states and the holelike FQH states
apart from the value of the filling fraction. In addition, Eq.\
(\ref{eq:bcs}) requires that the chiral bosons $\phi_I$ satisfy
the boundary conditions
\begin{equation}
\Delta \phi_I={\frac{2\pi}{2np+1}}
\left(
\begin{array}{ccc}
-p & 1  & 0\\
 1 & 2n & 0 \\
 0 & 0  & 2np+1
\end{array}
\right)_{IJ}
\left(
\begin{array}{c}
N_\phi \\
N_{qp} \\
-N_{qp}
\end{array}
\right)_J
\label{eq:bcs2}
\end{equation}
where $N_{qp}$ and $N_\phi$ are the total number of quasiparticles
(or charge) in the bulk and the extra magnetic flux in the bulk,
both with respect to the middle of the plateau, respectively.

\section{Electron and quasiparticle operators}
\label{sec:eops}

We will now seek a new basis of modes in which the quantum numbers
of the excitations are more transparent. We will use this representation
to construct the electron and quasiparticle operators at the edge.

Let us consider a generic operator that creates excitations, which can be
written as
\begin{equation}
\Psi (x) = e^{i (m_1 \phi_1 + m_2 \phi_2+m_3 \phi_3)}
\label{eq:opqp}
\end{equation}
The values that the coefficients $m_I$ take depend on the quantum
numbers of the particular quasiparticle that the operator  $\Psi
(x)$  creates. Recall that for physical states $m_2=-m_3$. It can
be shown (see for instance \cite{wen}) that these relations are
given by the following expressions
\begin{eqnarray}
     {\frac{Q} {e}}=&& \sum_{IJ} m_I K^{-1}_{IJ} t_J = {{-m_1 p +m_2}
\over {2np+1}}
\nonumber \\
{\theta \over \pi} =&&  \sum_{IJ} m_I K^{-1}_{IJ} m_J = - {{Q^2} \over
{\nu e^2}} + {{m_2^2}\over p}+m_2^2
\nonumber \\
&&
\label{eq:chst}
\end{eqnarray}
where $Q$ is the quasiparticle charge and $\theta$ is its statistics.

We have already identified the mode $\phi_1$ as the charge mode
and we will denote it as $\phi_1 \equiv \phi_C$. Eq.\
(\ref{eq:chst}) shows that an operator $\Psi$ with $m_2=p m_1$
creates neutral solitons. Although these states are not in general
part of the Hilbert space, we can nevertheless construct linear
combinations of the chiral bosons with these quantum numbers. We
will refer to these fields as the ``neutral modes". In particular,
it will be useful to rewrite the effective theory in terms of the
following linear combinations of the fields
\begin{eqnarray}
     \phi_C= &&\phi_1
\nonumber \\
      \phi_N =&& {\frac{1}{\sqrt p}} \phi_1 + {\sqrt p} \phi_2
\nonumber \\
     \phi_{N'}=&&\phi_3
\nonumber \\
&&
\label{eq:CN}
\end{eqnarray}
where we have introduced the ``neutral" modes $\phi_N$ and $\phi_{N'}$.

The edge effective Lagrangian  of Eq ~(\ref{eq:edge}) in terms of
the charged and neutral modes is diagonal,
\begin{eqnarray}
{\cal L}=&&-{\frac{1}{4\pi\nu}} \left(
 \partial_1\phi_C\partial_0\phi_C- v \partial_1\phi_C \partial_1\phi_C\right)
\nonumber \\
&&+{\frac{1}{4\pi}}\left(\partial_1\phi_N\partial_0 \phi_N+
\partial_1\phi_N\partial_0 \phi_{N'}
\right)
\label{eq:edgenew}
\end{eqnarray}
We see that only the charge mode $\phi_C$ propagates.
The role of the two remaining modes is to give the correct
quantum numbers to the quasiparticles, in particular their statistics.

The new fields $\phi_C$, $\phi_N$ and $\phi_{N'}$ obey the boundary conditions
\begin{eqnarray}
\Delta \phi_C=&&{\frac{2\pi}{2np+1}}\left(N_{qp}-p N_\phi\right)
\nonumber \\
\Delta \phi_N=&&{\frac{2\pi}{\sqrt{p}}} \; N_{qp}
\nonumber \\
\Delta \phi_{N'}=&&-2\pi N_{qp}
\nonumber \\
&&
\label{eq:bcsnew}
\end{eqnarray}

In the new basis of Eq.~( \ref{eq:CN}) the most general
quasiparticle operator of Eq.~(\ref{eq:opqp}) can be written as
\begin{equation}
\Psi (x) = e^{i (\alpha_C \phi_C + \alpha_N \phi_N+\alpha_{N'}\phi_{N'})}
\label{eq:opqp2}
\end{equation}
where
\begin{eqnarray}
\alpha_C=&& m_1-{\frac{m_2}{p}}
\nonumber \\
\alpha_N=&&{\frac{m_2}{\sqrt{p}}}
\nonumber \\
\alpha_{N'}=&&-m_2
\label{eq:relation}
\end{eqnarray}
Hence, the coefficients satisfy
\begin{eqnarray}
     {\frac{Q} {e}}=&& - \nu \alpha_C
\nonumber \\
{\theta \over \pi} =&& -\nu \alpha_C^2 +  \alpha_N^2 +\alpha_{N'}^2\\
&&
\label{eq:chst2}
\end{eqnarray}
where $Q$ is  the quasiparticle charge and $\theta$  its statistics.

The coefficients for the {\it quasiparticle} operator
should be such that they satisfy $Q=  {\frac {-e}{2np+1}}$ and
${\frac{\theta}{\pi}} ={\frac {2n}{2np+1}}+1$.
Therefore we find
\begin{equation}
 \alpha^{qp}_C=  {\frac {1}{p}} \; , \;  \alpha^{qp}_N =- {\frac {1}{\sqrt{p}}}
 \; , \; \alpha^{qp}_{N'}=1
\label{eq:qp11}
\end{equation}
which is consistent with setting $m_1=0$ and $m_2=-1$.

Likewise, to create an {\it electron} is be equivalent to create
$2np+1$ quasiparticles and hence it is defined by the choice
\begin{equation}
\alpha^e_C={\frac{1}{\nu}} \; , \; \alpha^e_N=
-{\frac{{\sqrt{p}}}{\nu}} \; , \; \alpha^{e}_{N'}={\frac{p}{\nu}}
\label{eq:eqns}
\end{equation}
It is immediate to show that this operator creates a state has
charge $Q=-e$ and statistics $\pi k$, where $k=(2np+1)[2n(p+1)+1]$
is an {\it odd} integer.

Thus, in this new basis, the {\it quasiparticle} operator is
\begin{equation}
\Psi_{qp}=e^{i({\frac{1}{p}}\phi_C-{\frac{1}{\sqrt{p}}}\phi_N+\phi_{N'})}
\label{eq:qp}
\end{equation}
and the {\it electron} operator has the form
\begin{equation}
\Psi_{e}=e^{i({\frac{1}{\nu}}\phi_C-{\frac{\sqrt{p}}{\nu}}\phi_N+
{\frac{p}{\nu}}\phi_{N'})}
\label{eq:eop}
\end{equation}
It is straightforward to show that if an integer number of {\it
electrons} $\Delta N_e$ is added to the bulk of the system, the
electron operator is not affected by the twist in the boundary
conditions Eq.~( \ref{eq:bcsnew}) since the exponent shifts by
$2\pi s $, where $s=[2n(p+1)+1](2np+1)\Delta N_e$ is an integer.

Finally, we will compute the propagators for the electron and the
quasiparticle operators. We will need the propagators of the
chiral bosons $\phi_C$, $\phi_N$ and $\phi_{N'}$. Since the
Lagrangians for $\phi_N$ and $\phi_{N'}$ are identical their
propagators are the same. Furthermore the chiral bosons  $\phi_N$
and $\phi_{N'}$ do not propagate ({\it i.\ e.\/} their velocity is
zero).

The propagator of the charged mode $\phi_C$, in imaginary time, is
\begin{eqnarray}
\langle \phi_C(x,t)  \phi_C(0,0) \rangle=&&
\nonumber \\
-{\frac{\nu}{2}}\ln &&
\left(1-{\frac{z^2}{\epsilon^2}}\right)+{\frac{\nu}{2}} {\rm sgn} (t)
\ln \left( {\frac{\epsilon+iz}{\epsilon-iz}}\right)
\nonumber \\
&&
\label{eq:gffic}
\end{eqnarray}
where $z=x+ivt$ and $\epsilon$ is an ultraviolet cutoff.
As $\epsilon \to 0$ we find
\begin{eqnarray}
\langle \phi_C(x,t)  \phi_C(0,0) \rangle
= &&
-\nu \ln {\frac {iz}{\epsilon}} +i \nu {\frac{\pi}{2}} {\rm sgn} (t)
\nonumber \\
&&
\label{eq:gffic2}
\end{eqnarray}
Notice that the (regulated) propagator obeys $\langle
\phi_C(0,0) \phi_C(0,0) \rangle=0$. The same applies to the
propagator of the neutral modes discussed below.

Likewise, the propagator (in imaginary time) of the neutral
modes $\phi_N$ and $\phi_{N'}$, in the same limit $\epsilon \to 0$, becomes
\begin{equation}
\langle \phi_N(x,t) \phi_N(0,0) \rangle=\langle \phi_{N'}(x,t) \phi_{N'}(0,0)
\rangle=-i{\frac{\pi}{2}} {\rm sgn}(t)
\label{eq:gffin}
\end{equation}

Using the propagators of Eq.\ (\ref{eq:gffic}) and Eq.\
(\ref{eq:gffin}) we find that the {\it electron} propagator is
given by \breakon
\begin{eqnarray}
\langle \Psi^\dagger_e(x,t) \Psi_e(0,0) \rangle =&&
\exp \left[{\frac{1}{\nu^2}} \langle \phi_C(x,t) \phi_C (0,0) \rangle )
+{\frac{p}{\nu^2}}\langle \phi_N(x,t) \phi_N(0,0) \rangle+{\frac{p^2}{\nu^2}}
\langle \phi_{N'}(x,t) \phi_{N'}(0,0) \rangle \right]
\nonumber \\
\to   && \equiv
{\frac{1}{|t|^{1/\nu}}}e^{i{\frac{\pi}{2}}({\frac{1}{\nu}-{\frac{p+p^2}{\nu^2}})
{\rm sgn}(t)} } = {\frac{1}{|t|^{1/\nu}}} e^{-i{\frac{\pi}{2}} {\rm sgn}(t)}
\label{eq:elprop}
\end{eqnarray}
\breakoff
where we have analytically continued to real time $t$ and taken the limit $x
\to 0$.

Eq.\ (\ref{eq:elprop}) shows clearly that the electron operator of
the Jain states with filling fraction $\nu$ has scaling dimension
$(2\nu)^{-1}$. This result implies that the tunneling density of
states for electrons at this edge obeys the law $|\omega|^
{(1-\nu)/\nu}$. Notice that the non-propagating modes are
responsible for the fermionic statistics of the electron.

Eq.\ (\ref{eq:elprop}) agrees with  the work by D.\ H.\ Lee and
X.\ G.\ Wen \cite{wenlee}, who have found independently the same
result as this paper was being written. However, in reference
\cite{wenlee} the neutral modes have a very different physical
origin and they result from considering the role of the
microscopic structure of the edge and edge reconstruction.
Instead, in the approach that we present in this paper the neutral
modes originate from global topological consistency requirements
for flux attachement and are a remnant of the topological
invariance of the Chern-Simons theory. Our result also agrees with
the recent work of U.\ Z{\"u}lucke and A.\ H.\
MacDonald\cite{alan} who calculated the electron tunneling
spectral function using a variational approach. These authors
found that although they could account for the correct spectral
function, their electron operator did not obey Fermi statistics,
in contrast with the result of Eq.\ (\ref{eq:elprop}).

The construction of the electron operator that we just derived
also has the following interesting interpretation. The electron
operator, as given by Eq.\ (\ref{eq:eop}), is a product of the
operator $\exp({\frac{i}{\nu}}\phi_C)$, which carries the charge,
and the operators $\exp(-{\frac{i}{\sqrt{p}}}\phi_N)$ and
$\exp(i\phi_{N'})$, which combined fix the statistics. In fact
this is the only role of these latter operators since the fields
$\Phi_N$ and $\phi_{N'}$ do not propagate. Essentially, the
combined operator $\exp(-{\frac{i}{\sqrt{p}}}\phi_N+i\phi_{N'})$
must be regarded as an effective Klein factor. In particular, it
also means that in a local probe of the edge, such as in electron
tunneling, only the charge mode plays a dynamical role. We will
discuss this problem elsewhere.

Finally, a similar calculation yields the {\it quasiparticle} propagator,
in imaginary time, which is found to be given by the following
expression
\breakon
\begin{eqnarray}
\langle \Psi_{qp}(x,t) \Psi_{qp}(0,0)\rangle
=&&
\exp \left[{\frac{1}{p^2}} \langle \phi_C(x,t) \phi_C (0,0) \rangle )
+{\frac{1}{p}}\langle \phi_N(x,t) \phi_N(0,0) \rangle+
\langle \phi_{N'}(x,t) \phi_{N'}(0,0) \rangle \right]
\nonumber \\
\to  &&
{\frac{1}{|t|^{\nu/p^2}}}e^{i{\frac{\pi}{2}}({\frac{1}{\nu}-{\frac{p+p^2}{\nu^2}})
{\rm sgn}(t)} }\equiv {\frac{1}{|t|^{\nu/p^2}}}
e^{-i{\frac{\theta_{qp}}{2}} {\rm sgn}(t)}
\label{eq:qpprop}
\end{eqnarray}
\breakoff again in the limit $x \to 0$. Eq.\ (\ref{eq:qpprop})
shows that the quasiparticle operator has scaling dimension
${\frac{\nu}{2p^2}}$ and the correct statistics.

\section{Conclusions}
\label{sec:concl}

In this work we have derived an  effective Chern-Simons theory for
the Jain states in a finite geometry that is  consistent with
global gauge invariance. We showed that this theory can be cast
into a $K$-matrix form but with a different and much simpler
structure than the usual one. We found that this structure
requires only a small number of gauge fields and their number,
that is the rank of the $K$ matrix, is fixed.
 We used this effective theory on a closed surface to find a
universal minimal structure of the theory on an open surface and
determine the structure of
 the edge states, for all the states in the
Jain sequences. We found that, in all cases, there is one and only
one propagating mode and hence only one mode that carries electric
charge and energy. The  role of the remaining (two) modes is to
fix the statistics of the excitations. We constructed the electron
and quasiparticle operators for these states, which turn out to be
uniquely determined and carry the correct charge and statistics.
We calculated the propagators of the excitations at the edge and,
in particular, found that the propagator for the electron (which
is a fermion as it should be) behaves as a function of time like
$|t|^{-1/\nu}$ for all the Jain states, and a tunneling density of
states that, as a function of frequency, behaves like
$|\omega|^{(1-\nu)/\nu}$. In a separate publication we generalize
these results to other quantum Hall effects\cite{other}.

These results in essence agree with the very recent work of D.\
H.\ Lee and X.\ G.\ Wen \cite{wenlee}. In particular, they also
find only one propagating mode which carries the charge current in
addition to a non-propagating mode that fixes the statistics.
However, the physical origin of this latter mode appears to be
quite different from the ones we find here. In the work of D.\ H.\
Lee and X.\ G.\ Wen, the non-propagating mode is one that survived
after much of the edge structure of the $K$ matrix theory has been
integrated out and this mode does not propagate in the sense that
their velocity is much smaller that the velocity of the charge
mode. In contrast, in the structure that we find here the
non-propagating modes have a topological origin and that is why
they do not propagate (or contribute to the specific heat of the
system).
 Z{\"u}licke and MacDonald\cite{alan}  have also
found recently the same result for the tunneling density of states, although
the electron operator they use does not have the correct Fermi
statistics. This result was actually anticipated by
Wen\cite{wen-edge} and by Kane and Fisher\cite{KF}, who noted
that, at the level of the effective theory of the edge states,
even though this result would follow if only the charge mode is
kept, the statistics of this electron operator is fermionic only
for Laughlin states.

The universal structure of this effective theory has a number of
potentially important implications precisely because its form does
not change dramatically from one Jain state to the next. Firstly,
it is reasonable to expect that a generalization of this theory is
likely to give a smooth dependence of the tunneling density of
states with the filling fraction for a continuous range of
magnetic fields, as suggested by the experiments of M.\ Grayson
and collaborators\cite{grayson}. However, to explain these
experiments is loosely equivalent to have a description of the
transition between plateaus as seen from the edge. Such a
description does not exist yet. Secondly, it may be necessary to
reexamine under this light the arguments that led to the phase
diagram of Kivelson, Lee and Zhang\cite{klz}, since the selection
rules for the transitions between plateaus are superficially
related, to an extent, to the number of edge states of nearby
plateaus\cite{comment}.

\section{Acknowledgements}

We thank C.\ Chamon, J.\ Jain, S.\ Kivelson and C.\ Nayak for
stimulating discussions, and particularly Dung Hai Lee for an early
communication of his work with Xiao-Gang Wen. This work begun during a visit
of EF to Universidad de La Plata, Argentina and to
Insituto Balseiro, Bariloche, Argentina.
EF is very grateful to Fidel Schaposnik for his hospitality in La
Plata, and to Andr{\'e}s Garc{\'\i}a and Manuel Fuentes for their
 hospitality in Bariloche. EF is a participant at
the ITP Program on {\it Disorder and Interactions in Quantum Hall
and Mesoscopic Systems}, and to D.\ Gross, Director of Institute
for Theoretical Physics of the University of California Santa
Barbara, for his kind hospitality. This work was supported in part
by the NSF grant number NSF DMR94-24511 at UIUC, and NSF
PHY94-07194 at ITP-UCSB (EF), and by CONICET (AL).

\end{multicols}

\end{document}